\documentclass[a4paper,11pt]{article}
\pdfoutput=1 %
\usepackage{jheppub} % 
\usepackage[T1]{fontenc} % if needed
\usepackage{comment}
\usepackage{booktabs}
\usepackage{subfig}
\usepackage{soul}
\usepackage{bm}
\usepackage{multirow}
\newcommand{\mysmall}[1]{\scriptscriptstyle \rm #1}
\newcommand{ \deltaw}{\delta_{\mysmall{W}}}
\newcommand{ \mw}{M_{\mysmall{W}}}
\newcommand{ \rw}{r_{\mysmall{W}}}
\newcommand{ \gw}{g_{\mysmall{W}}}

\newcommand{\ymin}{y_0}

\newcommand{\gmt}{$g$$-$$2$}

\newcommand{ \cw}{\cos \theta_{\mysmall W}}
\newcommand{ \sw}{\sin \theta_{\mysmall W}}

\newcommand{ \Clb}{C_{lB}}
\newcommand{ \Clw}{C_{lW}}

\renewcommand{\Re}{{\rm Re}}
\renewcommand{\Im}{{\rm Im}}
\title{\boldmath $\tau$ dipole moments via radiative leptonic $\tau$ decays}
\author[a,b]{S.~Eidelman}
\author[a,b,c]{D.~Epifanov}
\author[d]{M.~Fael}
\author[e]{L.~Mercolli}
\author[f]{M.~Passera}
\affiliation[a]{Budker Institute of Nuclear Physics SB RAS, Novosibirsk 630090, Russian Federation}
\affiliation[b]{Novosibirsk State University, Novosibirsk 630090, Russian Federation}
\affiliation[c]{The University of Tokyo, 7-3-1 Hongo Bunkyo-ku, Tokyo 113-0033, Japan}
\affiliation[d]{Albert Einstein Center for Fundamental Physics,\\
Institute for Theoretical Physics, University of Bern, CH-3012 Bern, Switzerland}
\affiliation[e]{Federal Office of Public Health FOPH, CH-3003 Bern, Switzerland}
\affiliation[f]{Istituto Nazionale Fisica Nucleare, Sezione di Padova, I-35131 Padova, Italy}
\emailAdd{eidelman@mail.cern.ch}
\emailAdd{d.a.epifanov@inp.nsk.su}
\emailAdd{fael@itp.unibe.ch}
\emailAdd{lorenzo.mercolli@bag.admin.ch}
\emailAdd{passera@pd.infn.it}

\begin{document} 
\abstract{We propose a new method to probe the magnetic and electric dipole moments of the $\tau$ lepton using precise measurements of the differential rates of radiative leptonic $\tau$ decays at high-luminosity $B$ factories. Possible deviations of these moments from the Standard Model values are analyzed in an effective Lagrangian approach, thus providing model-independent results. Analytic expressions for the relevant non-standard contributions to the differential decay rates are presented. Earlier proposals to probe the $\tau$ dipole moments are examined. A detailed feasibility study of our method is performed in the conditions of the Belle and Belle~II experiments at the KEKB and Super-KEKB colliders, respectively. This study shows that our approach, applied to the planned full set of Belle~II data for radiative leptonic $\tau$ decays, has the potential to improve the present experimental bound on the $\tau$ anomalous magnetic moment. On the contrary, its foreseen sensitivity is not expected to lower the current experimental limit on the $\tau$ electric dipole moment.}
%\keywords{Electromagnetic Processes and Properties}
\maketitle
\flushbottom

%%%%%%%%%%%%%%%%%%%%%%%%%%%%%%%%%%%%%%%%
\section{Introduction}\label{sec:intro}
%%%%%%%%%%%%%%%%%%%%%%%%%%%%%%%%%%%%%%%%

The very short lifetime of the $\tau$ lepton ($2.9 \times 10^{-13}$s) makes it very difficult to measure its electric and magnetic dipole moments.  While the Standard Model (SM) prediction of the $\tau$ anomalous magnetic moment $a_{\tau}=(g-2)_{\tau}/2$ is known with a tiny uncertainty of $5 \times 10^{-8}$~\cite{Eidelman:2007sb}, this short lifetime has so far prevented the determination of $a_{\tau}$ measuring the $\tau$ spin precession in a magnetic field, like in the electron and muon $g$$-$$2$ experiments. Instead, experiments focused on various high-precision measurements of $\tau$ pair production in high-energy processes, comparing the measured cross sections with the SM predictions. As these processes involve off-shell photons or taus in the $\tau \bar{\tau} \gamma$ vertices, the measured quantity is not directly $a_{\tau}$. The present resolution on $a_{\tau}$ obtained by these experiments is only of $O(10^{-2})$~\cite{Abdallah:2003xd}, more than an order of magnitude larger than its leading SM contribution $\frac{\alpha}{2\pi} \simeq 0.001$~\cite{Schwinger:1948iu}.

The electron and muon $g$$-$$2$, $a_e$ and $a_{\mu}$, have been measured with the remarkable precision of 0.24 ppb~\cite{Hanneke:2008tm} and $540$ ppb~\cite{Bennett:2006fi}, respectively. While $a_e$ perfectly agrees with the SM prediction~\cite{Aoyama:2012wj}, $a_{\mu}$, which is much more sensitive than $a_e$ to strong and weak interactions, shows a long-standing puzzling  discrepancy of about 3--4$\sigma$ and provides a powerful test of physics beyond the SM~\cite{Jegerlehner:2015stw, Jegerlehner:2009ry, Melnikov:2006sr, Passera:2004bj,Knecht:2003kc}. A precise measurement of $a_{\tau}$ would offer a new excellent opportunity to unveil new physics effects. Indeed, in a large class of theories beyond the SM, new contributions to the anomalous magnetic moment of a lepton $l$ of mass $m_l$ scale with $m_l^2$. Therefore, given the large factor $m_{\tau}^2/m_{\mu}^2 \sim 283$, the $g$$-$$2$ of the $\tau$ is much more sensitive than the muon one to electroweak and new physics loop effects that give contributions proportional to $m_l^2$. In these scenarios, the present discrepancy in the muon $g$$-$$2$ suggests a new-physics effect in $a_{\tau}$ of $\mathcal{O}(10^{-6})$; several theories exist where this naive scaling is violated and much larger effects are expected~\cite{Giudice:2012ms}.

The SM prediction of a lepton electric dipole moment (EDM) is extremely small and far below present experimental capabilities. Therefore, a measurement of a non-zero value would be direct evidence of new physics. Moreover, models for physics beyond the SM generally induce large contributions to lepton EDMs so that, although there has been no experimental evidence for an EDM so far, we hope that this kind of experiments will soon shed new light on the nature of $CP$ violation.

In this article we study the possibility to determine the electromagnetic dipole moments of the $\tau$ via the radiative leptonic decays $\tau \to l \gamma \nu \bar{\nu}$, with $l=\mu,e$, comparing the theoretical prediction for the differential decay rates with precise data from high-luminosity $B$ factories~\cite{Fael:2013ij,fael:phdthesis}. In particular, we present the results of a feasibility study performed in the conditions of the Belle~\cite{Abashian:2000cg,Brodzicka:2012jm,Akai:2001pf,Abe:2013kxa} and Belle~II~\cite{Abe:2010gxa} experiments at the KEKB~\cite{KEKB} and SuperKEKB~\cite{Aushev:2010bq,Ohnishi:2013fma} colliders, respectively. Following the strategy of the authors of refs.~\cite{GonzalezSprinberg:2000mk,Bernreuther:1993nd}, deviations of the $\tau$ dipole moments from the SM values are analyzed in an effective Lagrangian approach, thus avoiding the interpretation of off-shell form factors. We also examine the feasibility of earlier proposals; in particular, one based on the study of the Pauli form factor of the $\tau$ via $\tau^+ \tau^-$ production in $e^+ e^-$ collisions at the $\Upsilon$ resonances~\cite{Bernabeu:2007rr,Bernabeu:2008ii}, and another relying on the analysis of the radiation zero which occurs in radiative leptonic $\tau$ decays~\cite{Laursen:1983sm}.

In section~\ref{sec:ff} we establish our conventions for the $\tau$ electromagnetic form factors and introduce an effective Lagrangian to study the $\tau$ dipole moments. In section~\ref{sec:status} we review the present theoretical and experimental status on the $\tau$ \gmt~and EDM. The theoretical framework to analyze radiative leptonic $\tau$ decays is presented in section~\ref{sec:effectiveltau}, where we provide explicit analytic expressions for the relevant non-standard contributions to the differential decay rates. In section~\ref{sec:fstudy} we outline our method to determine the $\tau$ dipole moments and report the results of our feasibility study for the sensitivities that may be reached at the Belle and upcoming Belle~II experiments. Conclusions are drawn in sec.~\ref{sec:conclusions}.

%%%%%%%%%%%%%%%%%%%%%%%%%%%%%%%%%%%%%%%%
\section{\boldmath The $\tau$ lepton electromagnetic form factors} \label{sec:ff}
%%%%%%%%%%%%%%%%%%%%%%%%%%%%%%%%%%%%%%%%

Let us consider the structure of the $f{\bar f}\gamma$ coupling. The most general vertex function describing the interaction between a photon and the initial and final states of an arbitrary on-shell spin $1/2$ fermion $f$, with four-momenta $p$ and $p'$, respectively, can be written in the form
\begin{equation}
  \Gamma^\mu (q^2) = - i e Q_f \left\{
  \gamma^\mu F_1(q^2)   
  + \frac{\sigma^{\mu\nu} q_\nu}{2m_f} \Big[ i F_2(q^2) + F_3 (q^2) \gamma_5 \Big] 
  + \Big( \gamma^{\mu} - \frac{2q^{\mu}m_f}{q^2} \Big)  \gamma_5 \, F_4 (q^2) \right\},
  \label{eqn:ffgammavertex}
\end{equation}
where $e>0$ is the positron charge, $m_f$ is the mass of the fermion, $\sigma_{\mu\nu}=i/2\,[\gamma_\mu,\gamma_\nu]$, and $q=p'-p$ is the ingoing four-momentum of the off-shell photon. Equation~\eqref{eqn:ffgammavertex}, when sandwiched in $\overline{u}(p) \Gamma_\mu (q^2) u(p')$, is the most general expression that satisfies Lorentz and QED gauge invariance. The functions $F_{1}(q^2)$ and $F_{2}(q^2)$ are called the Dirac and Pauli form factors, respectively. In general, they are not physical quantities (for example, they can contain infrared divergences~\cite{Bonciani:2003ai,Mastrolia:2003yz}), but in the limit $q^2 \to 0$ they are measurable and related to the static quantities 
\begin{align}
  F_1(0) &= 1, & 
  F_2(0) &= a_f, &
  F_3(0) &= d_f \, \frac{2m_f}{e Q_f}, 
  \label{eqn:definitionformfactors} 
  \end{align}
where $e \, Q_f$ is the charge of the fermion, $a_f$ its anomalous magnetic moment, and $d_f$ its EDM. The electric dipole contribution $F_{3}(q^2)$ violates the discrete symmetries $P$ (parity) and $T$ (time reversal)~\cite{Barr:1988mc,Khriplovich:1997ga,Commins:1900zz}, and therefore $CP$, because of the $CPT$ theorem. $F_{4}(q^2)$ is called the anapole form factor and violates $P$.
In the limit $q^2 \to 0$, the dipole interactions in eq.~\eqref{eqn:ffgammavertex} can be cast in the form
\begin{equation}
   C_L \, \sigma_{\mu\nu} q^\nu P_L 
   + C_R \, \sigma_{\mu\nu} q^\nu P_R,
   \label{eqn:ffcomplex}
\end{equation}
where $P_{L,R} = (1 \mp \gamma^5)/2$. Hermiticity of this expression requires that $C_R = C_L^*=c_f$, with 
\begin{equation}
     c_f = a_f \frac{e Q_f}{2m_f} - i d_f, \quad \quad a_f, d_f \in \mathbb{R}.
     \label{eqn:cf}
\end{equation}  

Deviations of the $\tau$ dipole moments from the SM values can be analyzed in the framework of an effective field theory description where the SM Lagrangian is extended by a set of gauge-invariant higher-dimensional operators, built with the SM fields, suppressed by powers of the scale of new physics $\Lambda$~\cite{Buchmuller:1985jz}. We will consider only dimension-six operators, which are the lowest dimensional ones relevant for our analysis. Out of the complete set of 59 independent dimension-six operators in ref.~\cite{Grzadkowski:2010es}, only two of them can directly contribute to the $\tau$ lepton \gmt~and EDM at tree level (i.e., not through loop effects):
\begin{align}
  Q^{33}_{lW} &= \left( \bar{l}_\tau \sigma^{\mu\nu} \tau_R \right) \sigma^{\mysmall I} \varphi \,  W_{\mu\nu}^{\mysmall I}, 
  \label{eqn:QlW} \\
  Q^{33}_{lB} &= \left( \bar{l}_\tau \sigma^{\mu\nu} \tau_R \right) \varphi \, B_{\mu\nu}, \label{eqn:QlB}
\end{align}
where $\varphi$ and $l_\tau = (\nu_\tau,\tau_L)$ are the Higgs and the left-handed SU(2) doublets, $\sigma^{\mysmall I}$ are the Pauli matrices, and $W_{\mu\nu}^{\mysmall I}$ and $B_{\mu\nu}$ are the gauge field strength tensors. The leading non-standard effects will therefore arise from the effective Lagrangian
\begin{equation}
  \mathcal{L}_{\rm eff} =
  \frac{1}{\Lambda^2}
  \left[C^{33}_{lW}Q^{33}_{lW}+ 
  C^{33}_{lB} Q^{33}_{lB} + {\rm h.c.} \right].
  \label{eqn:leff}
\end{equation}
After the electroweak symmetry breaking, these two operators mix and give additional, beyond the SM, contributions to the $\tau$ anomalous magnetic moment and EDM:
\begin{align}
  \tilde{a}_\tau &= \frac{2 m_\tau}{e} \frac{\sqrt{2} v}{\Lambda^2} 
  \,\, \Re \left[ \cw \Clb^{33} -\sw \Clw^{33} \right] ,\\
  \tilde{d}_\tau &=  \frac{\sqrt{2} v}{\Lambda^2} 
  \,\, \Im \left[ \cw \Clb^{33} -\sw \Clw^{33} \right],
\end{align}
where $v=246$~GeV and $\sw$ is the weak mixing angle. Moreover, through the coupling to the $Z$ boson, the effective Lagrangian \eqref{eqn:leff} also gives non-standard contributions to the neutral weak dipole moments:
\begin{align}
  \tilde{a}_\tau^W &= \frac{2 m_\tau}{e} \frac{\sqrt{2} v}{\Lambda^2} 
   \,\, \Re \left[ \sw \Clb^{33} + \cw \Clw^{33} \right] ,\\
  \tilde{d}_\tau^W &=  -\frac{\sqrt{2} v}{\Lambda^2} 
  \,\, \Im \left[ \sw \Clb^{33} + \cw \Clw^{33} \right].
\end{align}

The operator $Q^{33}_{lW}$ in~\eqref{eqn:QlW} also generates an additional chirality-flipping coupling between the $\tau$ and the $W$ boson, and a four-point vertex that couples the $\tau$ and the $W$ to the photon or the $Z$ (other four- and five-point vertices, involving the physical Higgs boson, will not be considered since they do not contribute to the $\tau$ dipole moments nor to the decays $\tau \to l \nu \bar{\nu} (\gamma)$). These additional $\tau$-$W$ couplings are proportional to the complex parameter $\Clw^{33}$ and, therefore, to the real combinations 
$\tilde{b}_\tau = -(2m_{\tau}/e)(\sqrt{2}v/\Lambda^2) \sw \, \Re \, \Clw^{33} = \sin^2 \! \theta_{\mysmall W} \tilde{a}_\tau - \sw \cw \tilde{a}_\tau^W$
and 
$\tilde{c}_\tau = -(\sqrt{2}v/\Lambda^2) \sw \, \Im \, \Clw^{33} = \sin^2 \! \theta_{\mysmall W} \tilde{d}_\tau + \sw \cw \tilde{d}_\tau^W$.
The dynamics of radiative leptonic $\tau$ decays is modified both by non-standard terms proportional to $\tilde{a}_\tau$ and $\tilde{d}_\tau$ (see section~\ref{sec:effectiveltau}), as well as by contributions generated by these new couplings between the $\tau$ and the $W$ boson, which are proportional to $\tilde{b}_\tau$ and $\tilde{c}_\tau$. However, as these new $\tau$-$W$ couplings also affect the ordinary (inclusive) leptonic $\tau$ decays $\tau \to l \nu \bar{\nu}$, we will assume that future bounds on $\tilde{b}_\tau$ and $\tilde{c}_\tau$ will be more stringent than those on $\tilde{a}_\tau$ and $\tilde{d}_\tau$ obtained via radiative leptonic decays. The present limits on $\tilde{b}_\tau$ and $\tilde{c}_\tau$ are of $\mathcal{O}(10^{-3})$; should future bounds on $\tilde{a}_\tau$ and $\tilde{d}_\tau$ reach the sensitivity of $\tilde{b}_\tau$ and $\tilde{c}_\tau$, then a combined analysis of ordinary and radiative leptonic $\tau$ decays for $\tau$ dipole moments and Bouchiat-Michel-Kinoshita-Sirlin parameters~\cite{Michel:1949qe,Bouchiat:1957zz,Kinoshita:1957zz,Kinoshita:1957zza} will become necessary. For the time being, we will neglect these new $\tau$-$W$ couplings.

%%%%%%%%%%%%%%%%%%%%%%%%%%%%%%%%%%%%%%%%
\section{{\boldmath Status of the $\tau$ lepton $g$-2 and EDM}}    \label{sec:status}
%%%%%%%%%%%%%%%%%%%%%%%%%%%%%%%%%%%%%%%%

In this section we discuss the present status of the SM prediction and experimental determination of the anomalous magnetic moment and EDM of the $\tau$ lepton.

%% g-2 TH ----------------------------------------------------------

The SM prediction for $a_{\tau}$ is given by the sum of QED, electroweak (EW) and hadronic terms. The QED contribution has been computed up to three loops:
$
    a_{\tau}^{\mysmall \rm QED} =
    117 \, 324 \, (2) \times 10^{-8}
$~\cite{Laporta:1992pa,Laporta:1993ju,Laporta:1996mq,Passera:2006gc}, where the uncertainty
$\pi^2 \ln^2(m_{\tau}/m_e)(\alpha/\pi)^4 \sim 2\times 10^{-8}$
has been assigned for uncalculated four-loop contributions. The errors due to the uncertainties of the $\mathcal{O}(\alpha^2)$ and $\mathcal{O}(\alpha^3)$ terms, as well as that induced by the uncertainty of $\alpha$, are negligible.
The sum of the one- and two-loop EW contributions is
$
    a_{\tau}^{\mysmall \rm EW} = 47.4 (5) \times 10^{-8}
$~\cite{Czarnecki:1995wq,Czarnecki:1995sz,Eidelman:2007sb}. The uncertainty encompasses the estimated errors induced by hadronic loop effects, neglected two-loop bosonic terms and the missing three-loop contribution. It also includes the tiny errors due to the uncertainties in $m_{\rm\scriptstyle top}$ and $m_{\tau}$.

Similarly to the case of the muon $g$$-$$2$, the leading-order hadronic contribution to $a_{\tau}$ is obtained via a dispersion integral of the total hadronic cross section of the $e^+e^-$ annihilation (the role of low energies is very important, although not as much as for $a_{\mu}$). The result of the latest evaluation, using experimental data below 12~GeV, is
$
    a_{\tau}^{\mysmall \rm HLO} =  337.5 \, (3.7) \times 10^{-8}
$~\cite{Eidelman:2007sb}.
The hadronic higher-order $(\alpha^3)$ contribution $a_{\tau}^{\mysmall \rm HHO}$ can be divided into two parts:
$
     a_{\tau}^{\mysmall \rm HHO}=
     a_{\tau}^{\mysmall \rm HHO}(\mbox{vp})+
     a_{\tau}^{\mysmall \rm HHO}(\mbox{lbl}).
$
The first one, the $\mathcal{O}(\alpha^3)$ contribution of diagrams containing hadronic self-energy insertions in the photon propagators,
is
$
a_{\tau}^{\mysmall \rm HHO}(\mbox{vp})= 7.6 (2) \times 10^{-8}
$~\cite{Krause:1996rf}.
Note that naively rescaling the corresponding muon $g$$-$$2$ result by a factor $m_{\tau}^2/m_{\mu}^2$ leads to the incorrect estimate $a_{\tau}^{\mysmall \rm HHO}(\mbox{vp}) \sim -28\times 10^{-8}$ (even the sign is wrong!). 
Estimates of the light-by-light contribution $a_{\tau}^{\mbox{$\scriptscriptstyle{\rm HHO}$}}(\mbox{lbl})$ obtained rescaling the corresponding one for the muon $g$$-$$2$ by a factor $m_{\tau}^2/m_{\mu}^2$ fall short of what is needed -- this scaling is not justified. The parton-level estimate of~\cite{Eidelman:2007sb} is
$
a_{\tau}^{\mysmall \rm HHO}(\mbox{lbl})= 5 (3) \times 10^{-8},
$
a value much lower than those obtained by naive rescaling. Adding up the above contributions one obtains the SM 
prediction~\cite{Eidelman:2007sb}
\begin{equation}
    a_{\tau}^{\mysmall \rm SM} = 
         a_{\tau}^{\mysmall \rm QED} +
         a_{\tau}^{\mysmall \rm EW}  +
         a_{\tau}^{\mysmall \rm HLO}  +
         a_{\tau}^{\mysmall \rm HHO}
         =117 \, 721 \, (5) \times 10^{-8}.  
\label{eqn:atSM}
\end{equation}
Errors were added in quadrature.

%% EDM TH 	----------------------------------------------------------

The EDM interaction violates the discrete $CP$ symmetry. In the SM with massless neutrinos, the only source of $CP$ violation is the CKM-phase (and a possible $\theta$-term in the QCD sector). In refs.~\cite{Jarlskog:1985cw,Jarlskog:1985ht} it was shown that all $CP$-violating amplitudes are proportional to the Jarlskog invariant $J$, defined as
\begin{equation}
   \text{Im} \left[ V_{ij} V_{kl} V^*_{il} V^*_{kj} \right] =
   J \sum_{m,n} \varepsilon_{ikm} \varepsilon_{jln} \, ,
\end{equation}
where $V_{ij}$ are the CKM matrix elements. Therefore, the lepton EDM must arise from virtual quarks linked to the lepton through the $W$ boson, thus being sensitive to the imaginary part of the CKM matrix elements.
The leading contribution is naively expected at the three-loop level, since two-loop diagrams are proportional to $|V_{ij}|^2$. The problem was first analyzed in some detail in~\cite{Hoogeveen:1990cb}, but it was subsequently shown that also three-loop diagrams yield a zero EDM contribution in the absence of gluonic corrections to the quark lines~\cite{Pospelov:1991zt}. For this reason, lepton EDMs are predicted to be extremely small in the SM, of the $\mathcal{O}(10^{-38} - 10^{-35}) \, e\cdot$cm~\cite{Commins:1900zz}, far below the present $\mathcal{O}(10^{-17}) \, e\cdot$cm experimental reach on the $\tau$ EDM. Even for the electron, the fantastic experimental upper bound $d_e^{\mysmall EXP} < 0.87 \times 10^{-28} ~e\cdot$cm~\cite{Baron:2013eja} is still much larger than the SM prediction $d_e^{\mysmall SM} \sim \mathcal{O}(10^{-38}) \, e \cdot $cm and it is hard to imagine improvements in the sensitivity by ten orders of magnitude! However, new EDM effects could arise at the one- or two-loop level from new physics that violates $P$ and $T$, and be much larger than the tiny SM value, even if they arise from high mass scales.

% g-2 EXP	----------------------------------------------------------
The present experimental resolution on the $\tau$ anomalous magnetic moment is only of $\mathcal{O}(10^{-2})$~\cite{Abdallah:2003xd}, more than an order of magnitude larger than its SM prediction in Eq.~\eqref{eqn:atSM}. In fact, while the SM value of $a_{\tau}$ is known with a tiny uncertainty of $5 \times 10^{-8}$, the $\tau$ short lifetime has so far prevented the determination of $a_{\tau}$ by measuring the $\tau$ spin precession in a magnetic field, like in the electron and muon $g$$-$$2$ experiments. 
The present PDG limit on the $\tau$ \gmt~was derived in 2004 by the DELPHI collaboration from $e^+ e^- \to e^+ e^- \tau^+ \tau^-$ total cross section measurements at $\sqrt{s}$  between 183 and 208 GeV at LEP2 (the study of $a_\tau$ via this channel was proposed in~\cite{Cornet:1995pw}). The measured values of the cross-sections were used to extract limits on the $\tau$ \gmt~by comparing them to the SM values, assuming that possible deviations were due to non-standard contributions $\tilde{a}_\tau$. The obtained limit at 95\%~CL is~\cite{Abdallah:2003xd}
\begin{equation}
   -0.052 < \tilde{a}_\tau < 0.013,
   \label{eqn:atauexpbound95}
\end{equation}
which can be also expressed in the form of central value and error as~\cite{Abdallah:2003xd}
\begin{equation}
   \tilde{a}_\tau = -0.018 \, (17).
   \label{eqn:atauexpbound68}
\end{equation}
%

%% EDM	EXP ----------------------------------------------------------
The present PDG limit on the EDM of the $\tau$ lepton at $95\%$~CL is
\begin{equation}\label{eq dtauexp}
\begin{split}
& - 2.2 < \mathrm{Re} (d_\tau) < 4.5 \; \; (10^{-17} \;  e \mathrm{\cdot cm}), \\
& - 2.5 < \mathrm{Im} (d_\tau) < 0.8 \; \; (10^{-17} \;  e  \mathrm{\cdot cm}); \\
\end{split}
\end{equation}
it was obtained by the Belle collaboration~\cite{Inami:2002ah} following the analysis of ref.~\cite{Bernreuther:1993nd} for the impact of an effective operator for the $\tau$ EDM in the process $e^+ e^- \rightarrow \tau^+ \tau^-$.

% indirect bound from LEP data 
The reanalysis of ref.~\cite{GonzalezSprinberg:2000mk} of various LEP and SLD measurements -- mainly of the $e^+e^- \to \tau^+\tau^-$ cross sections -- allowed the authors to set the indirect 2$\sigma$ confidence interval
\begin{equation}
   -0.007 < \tilde{a}_{\tau}  < 0.005, 
\end{equation}
a bound stronger than that in Eq.~(\ref{eqn:atauexpbound95}). This analysis assumed $\tilde{d}_\tau = 0$. We updated this analysis using more recent data~\cite{Schael:2013ita,Agashe:2014kda} obtaining the almost identical $2\sigma$ confidence interval $-0.007 < \tilde{a}_{\tau}  < 0.004$.

% LHC
At the LHC, bounds on the $\tau$ dipole moments are expected to be set in $\tau$ pair production via Drell-Yan~\cite{Hayreter:2013vna,Hayreter:2015cia} or double photon scattering processes~\cite{Atag:2010ja}. The best limits achievable in $pp \to \tau^+\tau^- + X$ are estimated to be comparable to present existing ones if the total cross section for $\tau$ pair production is assumed to be measured at the $14\%$ level~\cite{Hayreter:2013vna}. Earlier proposals to set bounds on the $\tau$ dipole moments can be found in~\cite{delAguila:1991rm,Samuel:1992fm,Escribano:1993pq,Escribano:1996wp}.

Yet another method to determine $\tilde{a}_{\tau}$ would use the channeling of polarized $\tau$ leptons in a bent crystal similarly to the suggestion for the measurement of magnetic moments of short-living baryons~\cite{Kim:1982ry}. This approach has been successfully tested by the E761 collaboration at Fermilab, which measured the magnetic moment of the $\Sigma^+$ hyperon~\cite{Chen:1992wx}. The challenge of this method is to produce a polarized beam of $\tau$ leptons. One could use the decay $B^+ \to \tau^+ \nu_\tau$, which would produce polarized $\tau$ leptons~\cite{Samuel:1990su}; however this particular decay of the $B$ has a very tiny branching ratio of $\mathcal{O} ( 10^{-4})$. In 1991, when this proposal was published, the idea seemed completely unlikely. Nonetheless, in the era of $B$ factories, when the decay $B^+ \to \tau^+ \nu_\tau$ is already observed~\cite{Agashe:2014kda}, the realization of this idea in a dedicated experiment is definitively not excluded.

The Belle II experiment at the upcoming high-luminosity $B$ factory SuperKEKB will offer new opportunities to improve the determination of the $\tau$ electromagnetic properties. The authors of ref.~\cite{Bernabeu:2007rr,Bernabeu:2008ii} proposed to determine the Pauli form factor $F_{2}(q^2)$ of the $\tau$ via $\tau^+ \tau^-$ production in $e^+ e^-$ collisions at the $\Upsilon$ resonances ($\Upsilon$(1S), $\Upsilon$(2S) and $\Upsilon$(3S)) with a sensitivity of $\mathcal{O}(10^{-5})$ or even better (of course, the center-of-mass energy at super $B$ factories is $\sqrt{s} \sim M_{\Upsilon(4S)} \approx 10$ GeV, so that the form factor $F_{2}(q^2)$ is not the anomalous magnetic moment). When attempting to extract the value of $F_{2}(q^2)$ from scattering experiments (as opposed to using a background magnetic field) one encounters additional complications due to the contributions of various other Feynman diagrams not related to the magnetic form factor. In particular, in the $e^+e^- \to \tau^+ \tau^-$ case, contributions to the cross section arise not only from the usual $s$-channel one-loop vertex corrections, but also from box diagrams, which should be somehow subtracted out. The strategy proposed in~\cite{Bernabeu:2007rr,Bernabeu:2008ii} to eliminate their contamination is to measure the observables on top of the $\Upsilon$ resonances, where the non-resonant box diagrams should be numerically negligible.

However, because of the natural irreducible beam energy spread associated to any $e^+ e^-$ synchrotron, it is very difficult to resolve the narrow peaks of the $\Upsilon (1S,2S,3S)$ in the $\tau^+ \tau^-$ decay channel (the $\Upsilon(4S)$ decays almost entirely in $B\bar{B}$). Indeed, the total visible cross section of these resonances is not a perfect Breit-Wigner, but the convolution of the theoretical Breit-Wigner cross section with a Gaussian spread,
\begin{equation}
   \sigma_{\mysmall vis} = 
   \int \frac{\sigma_{ee\to\Upsilon\to\tau\tau}(s)}{\sqrt{2 \pi} \sigma_{\mysmall W}} 
   \, \exp \! \left[ - \frac{(\sqrt{s}-M_\Upsilon)^2}{2 \sigma_{\mysmall W}^2} \right]
   d\sqrt{s},
   \label{eqn:visxsec}
\end{equation}
where $\sigma_{\mysmall W}$ is the irreducible beam energy spread of the accelerator at $\sqrt{s} = M_\Upsilon$ ($\sigma_{\mysmall W}=5.45$~MeV at the upcoming SuperKEKB collider), $\sigma_{ee\to\Upsilon\to\tau\tau}(s)$ is the total cross section in the Breit-Wigner approximation,
\begin{equation}
   \sigma_{ee\to\Upsilon\to\tau\tau}(s) \, = \, \sigma_{\mysmall peak} \, 
   \frac{M_\Upsilon^2\Gamma_{\Upsilon}^2}{(s-M_\Upsilon^2)^2 + M_\Upsilon^2\Gamma_{\Upsilon}^2}, 
   \label{eqn:peakxsec}
\end{equation}
$M_\Upsilon$ and $\Gamma_{\Upsilon}$ are the masses and the widths of the $\Upsilon$ resonances, and the cross section at the peak is  given by $\sigma_{\mysmall peak} = 12 \pi {\cal B} ({\Upsilon \to ee}) {\cal B} ({\Upsilon \to \tau\tau})/M_\Upsilon^2$. In the limit $\Gamma_{\Upsilon} \ll \sigma_{\mysmall W}$ of narrow resonances, $\sigma_{ee\to\Upsilon\to\tau\tau}(s)$ can be approximated by 
\begin{equation}
    \sigma_{ee\to\Upsilon\to\tau\tau}(s) \approx 
    \sigma_{\mysmall peak} \pi M_\Upsilon \Gamma_{\Upsilon} \delta (s-M_\Upsilon^2).    
\end{equation}
The expression for the maximum visible resonant cross section obtained substituting eq.~\eqref{eqn:peakxsec} into eq.~\eqref{eqn:visxsec} is
\begin{equation}
   \sigma_{\mysmall vis}^{\mysmall max} = \rho \, \sigma_{\mysmall peak}, \quad \mbox{with} \quad 
   \rho = \sqrt{\frac{\pi}{8}} \, \frac{\Gamma_{\Upsilon}}{\sigma_{\mysmall W}}.
   \label{eqn:finalvisxsec}
\end{equation}
In table~\ref{tab:ures} we compare the maximum visible resonant cross sections for $e^+ e^- \to \Upsilon \to \tau^+ \tau^-$ with the non-resonant cross section $\sigma_{\mysmall non-res} = 0.919(3)$~nb at $\sqrt{s} = M_\Upsilon$~\cite{Banerjee:2007is}. From this table we can conclude that, at the Belle II experiment, the $\tau^+ \tau^-$ events produced with beams at a center-of-mass energy $\sqrt{s} \sim M_\Upsilon$ are mostly due to non-resonant contributions; indeed the visible resonant cross sections are of the same order of the non-resonant ones, or smaller. Even for the multihadron events in the region of $\Upsilon(1S,2S,3S$), the non-resonant cross section dominates with respect to the resonant one (see, for example,~\cite{Artamonov:1983vz}). The situation at Belle was similar (the energy spread at KEKB was $\sigma_{\mysmall W} = 5.24$ MeV~\cite{KEKB}). We therefore conclude that measuring the $e^+ e^- \to \tau^+ \tau^-$ cross section at the upcoming SuperKEKB collider on top of the $\Upsilon$ resonances will not eliminate the contamination of the non-resonant contributions. 
\begin{table}
   \centering
   \begin{tabular}{l|c|c|c|c|c}
      \toprule
      $\Upsilon$ & $M_\Upsilon$ [GeV] & $\Gamma_{\Upsilon}$ [keV] & 
      $\sigma_{\mysmall peak}$ [nb] & $\rho$ & $\displaystyle  \frac{\sigma_{\mysmall vis}^{\mysmall max}}{\sigma_{\mysmall non-res}}$ \\
      \midrule
      $\Upsilon(1S)$ & $\phantom{1}9.46$ 	& $54$ & $101$ & $6.2 \times 10^{-3}$ & $69\%$ \\
      $\Upsilon(2S)$ & $10.02$ 			& $32$ & $56$   & $3.7 \times 10^{-3}$ & $22\%$ \\
      $\Upsilon(3S)$ & $10.36$ 			& $20$ & $68$   & $2.3 \times 10^{-3}$ & $17\%$ \\
      $\Upsilon(4S)$ & $10.58$ 			& $20 \times 10^3$ & -- & -- & -- \\
      \bottomrule
   \end{tabular}
   \caption{Estimated visible cross section at Belle II for $e^+ e^- \to \Upsilon \to \tau^+ \tau^-$.
   The machine parameters are from ref.~\cite{Ohnishi:2013fma}.}
   \label{tab:ures}
\end{table}

In the next section we will propose a new method to determine the electromagnetic dipole moments of the $\tau$ lepton via precise measurements of its radiative leptonic decays.

%%%%%%%%%%%%%%%%%%%%%%%%%%%%%%%%%%%%%%%%
\section{{\boldmath Radiative $\tau$ leptonic decays: theoretical framework}} \label{sec:effectiveltau}
%%%%%%%%%%%%%%%%%%%%%%%%%%%%%%%%%%%%%%%%

The SM prediction, at next-to-leading order (NLO), for the differential rate of the radiative leptonic decays 
\begin{equation}
   \tau^- \to l^- \, \nu_\tau \, \bar{\nu}_l \, \gamma, 
   \quad 
   \label{eqn:raddecay}
\end{equation}
with $l=e$ or $\mu$, of a polarized $\tau^-$ with mass $m_{\tau}$ in its rest frame is
\begin{equation}
   \frac{d^6 \Gamma \left(\ymin\right) }{dx \, dy \, d\Omega_l\, d\Omega_\gamma}  =
	\frac{\alpha \, G_F^2 m_{\tau}^5} {(4 \pi)^6} 
	\frac{x \beta_l}{1+ \deltaw}
	\biggl[
	G
	\, + \, x \beta_l \, \hat{n} \cdot \hat{p}_l  \, J 
	\, + \, y \, \hat{n} \cdot \hat{p}_\gamma \, K 
        \, + \, x y \beta_l \, \hat{n} \cdot \left(\hat{p}_l \times \hat{p}_\gamma \right) L
	\biggr],
  \label{eqn:radiativedecayrateNLO}
\end{equation}
where 
$G_F=1.166 \, 378 \, 7(6) \times10^{-5}$ GeV$^{-2}$~\cite{Webber:2010zf} 
is the Fermi constant determined by the muon lifetime and
$\alpha = 1/137.035\,999\,157\,(33)$
is the fine-structure constant~\cite{Aoyama:2012wj,Aoyama:2014sxa}. 
Calling $m$ the mass of the final charged lepton (neutrinos and antineutrinos are considered massless) we define $r=m/m_{\tau}$ and $\rw=m_{\tau}/\mw$, where $\mw$ is the $W$-boson mass; $p$ and $n=(0,\hat{n})$ are the four-momentum and polarization vector of the initial $\tau$, with $n^2=-1$ and $n \cdot p = 0$. Also, $x = 2E_l/m_{\tau}$, $y = 2E_\gamma/m_{\tau}$ and $\beta_l \equiv |\vec{p}_l|/E_l=\sqrt{1-4r^2/x^2}$, where  $p_l = (E_l,\vec{p}_l)$ and $p_\gamma = (E_\gamma,\vec{p}_\gamma)$ are the four-momenta of the final charged lepton and photon, respectively. The final charged lepton and photon are emitted at solid angles $\Omega_l$ and $\Omega_{\gamma}$, with normalized three-momenta $\hat{p}_l$ and $\hat{p}_\gamma$, and  $c$ is the cosine of the angle between $\hat{p}_l$ and $\hat{p}_\gamma$. The term $ \deltaw =1.04 \times 10^{-6}$ is the tree-level correction to muon decay induced by the $W$-boson propagator~\cite{Ferroglia:2013dga,Fael:2013pja}.

Equation~\eqref{eqn:radiativedecayrateNLO} includes the possible emission of an additional soft photon with normalized energy $y'$ lower than the photon detection threshold $\ymin$ (with $\ymin \ll 1$): $y'<\ymin<y$.
The function $G (x,y,c,\ymin)$ and, analogously, $J$ and $K$, are given by
\begin{equation}
  G \, (x,y,c,\ymin) =
  \frac{4}{3 y z^2} 
  \left[ 
     g_0 (x,y,z) 
     + \rw^2 \, \gw  (x,y,z) 
     + \frac{\alpha}{\pi} \, g_{\mysmall NLO} (x,y,z,\ymin) 
   \right],
  \label{eqn:GNLO}
\end{equation}
where $z=xy(1-c\beta_l)/2$; the LO function $g_0 (x,y,z)$, computed in~\cite{Kinoshita:1958ru,Fronsdal:1959zzb,EcksteinPratt,Kuno:1999jp}, arises from the pure Fermi $V$--$A$ interaction, whereas $\gw(x,y,z)$ is the LO contribution of the $W$-boson propagator derived in~\cite{Fael:2013pja}. The NLO term $g_{\mysmall NLO} (x,y,z,\ymin)$ is the sum of the virtual and soft bremsstrahlung contributions calculated in~\cite{Fael:2015gua} (see also refs.~\cite{Fischer:1994pn,Arbuzov:2004wr}). The function $L(x,y,z)$, appearing in front of the product $\hat{n} \cdot \left(\hat{p}_l \times \hat{p}_\gamma \right)$, does not depend on $\ymin$; it is only induced by the loop corrections and is therefore of $\mathcal{O}(\alpha/\pi)$. In particular, $L(x,y,z)$ is of the form $\sum_n P_n(x,y,z) \,  {\rm Im} \left[I_n (x,y,z)\right]$, where $P_n$ are polynomials in $x,y,z$ and $I_n (x,y,z)$ are scalar one-loop integrals whose imaginary parts are different from zero. Tiny terms of $\mathcal{O}(\alpha \, m_{\tau}^2/\mw^2) \sim 10^{-6}$ were neglected; they are expected to be comparable to the uncomputed next-to-next-to-leading order (NNLO) corrections of $\mathcal{O}((\alpha/\pi)^2)$. The functions $G$, $J$, $K$ and $L$ are free of UV and IR divergences. Their (lengthy) explicit expressions are provided in~\cite{Fael:2015gua}. The corresponding formula for the radiative decay of a polarized $\tau^+$ can be simply obtained replacing $J \to -J$ and $K \to -K$ in eq.~\eqref{eqn:radiativedecayrateNLO} (see table~\ref{tab:tauplus}). If the initial $\tau^{\pm}$ are not polarized, eq.~\eqref{eqn:radiativedecayrateNLO} simplifies to
\begin{equation}	
\frac{d^3 \Gamma  \left(\ymin\right) }{dx \, dc \, dy}  =
	\frac{\,\alpha G_F^2 m_{\tau}^5} {(4 \pi)^6} \frac{x \beta_l}{1+ \deltaw}  \,\, 8 \pi^2 \, G \, (x,y,c,\ymin).
\label{eq:radiativedecayrateunpolarizedNLO}
\end{equation}
For the differential rate of leptonic $\tau$ decays in which a virtual photon is emitted and converted into a lepton pair, we refer the reader to the recent comprehensive article in~\cite{Flores-Tlalpa:2015vga}.

The effective Lagrangian \eqref{eqn:leff} generates additional non-standard contributions to the differential decay rate of a polarized $\tau^-$ in eq.~\eqref{eqn:radiativedecayrateNLO}.\footnote{As discussed in section~\ref{sec:ff}, we neglect non-standard $\tau$-$W$ couplings arising from the operator $Q^{33}_{lW}$.} They can be summarised in the shifts:
\begin{align}
	G   & \,\to\,   G  \,+\,  \tilde{a}_\tau \,  G_a, 
	\label{eqn:Gachange}\\
	J    & \,\to\,   J  \,+\,  \tilde{a}_\tau \,  J_a, 
	\label{eqn:Jachange}\\
	K   & \,\to\,   K  \,+\,  \tilde{a}_\tau \,  K_a,
	\label{eqn:Kachange}\\
	L    & \,\to\,   L  \,+\,  \left(m_\tau/e \right) \,  \tilde{d}_\tau \, L_d,
	\label{eqn:Ldchange}
\end{align}
where
\begin{align}
   G_a &=
   \frac{4}{3z}
   \left[r^2 \left(y^2-y z+3 z^2\right)-z (y+2 z) (x+y-z-1) \right],\\
%-----------------------
   J_a &=
   \frac{2}{3z}
   \big[
   3 r^2 \left(x y+y^2-2 z\right)-2 x^2 y-4 x y^2+2 x y z+x y +4 x z-2 y^3+2 y^2 z \notag \\
   & +2 y^2+3 y z-4 z^2-2 z \big], \\
%--------------------------
   K_a &= 
   \frac{2}{3 y z}
   \big[
   12 r^4 y+r^2 \left(-3 x^2 y-3 x y^2-8 x y-6 y^2+8 y z+4 y+6 z^2\right)
   +2 x^3 y+4 x^2 y^2\notag \\
   &-2 x^2 y z-x^2 y+2 x y^3-2 x y^2 z-2 x y^2-x y z -4 x z^2-2 y^2 z-2 y z^2+2 y z+4 z^3+2 z^2\big],\\
%-----------------------
 L_d &= 
 \frac{4}{3yz}
 \big[
   3 r^2 \left(x y+y^2-2 z\right)-2 x^2 y-4 x y^2+2 x y z+x y+4 x z-2 y^3 +2 y^2 z \notag \\ 
   &+2 y^2+3 y z-4 z^2-2 z
 \big]
\end{align}
(we note that $L_d=2J_a/y$). Tiny terms of $\mathcal{O}(\tilde{a}_{\tau}^2)$, $\mathcal{O}(\tilde{d_{\tau}}^2)$ and $\mathcal{O}(\tilde{a}_{\tau} \tilde{d_{\tau}})$ were neglected. For $\tau^+$ decays, the theoretical prediction for the differential decay rate can again be obtained from eq.~\eqref{eqn:radiativedecayrateNLO}, simply performing the following substitutions (see table~\ref{tab:tauplus}):
\begin{align}
	G   & \,\to\,   G  \,+\,  \tilde{a}_\tau \,  G_a, 
	\label{eqn:Gachangeplus}\\
	J   & \,\to\,   -J  \,-\,  \tilde{a}_\tau \,  J_a,
	\label{eqn:Jachangeplus}\\
	K   & \,\to\,   -K  \,-\,  \tilde{a}_\tau \,  K_a,
	\label{eqn:Kachangeplus}\\
	L   & \,\to\,   L  \,-\,  \left(m_\tau/e \right) \,  \tilde{d}_\tau \, L_d.
	\label{eqn:Ldchangeplus}
\end{align}
Deviations of the $\tau$ dipole moments from the SM values can be determined comparing the SM prediction for the differential rate in eq.~\eqref{eqn:radiativedecayrateNLO}, modified by the terms $G_a$, $J_a$, $K_a$ and $L_d$, with sufficiently precise data.

\begin{table}
   \centering
   \begin{tabular}{c||c|c|c|c|c|c|c|c}
      \toprule
      $\tau^-$ 	& $+G$ & $+J$ & $+K$ & $+L$ & $+G_a$ & $+J_a$ & $+K_a$ & $+L_d$\\
      \hline
      $\tau^+$ 	& $+G$ & $-J$ & $-K$ & $+L$ & $+G_a$ & $-J_a$ & $-K_a$ & $-L_d$\\
      \bottomrule
   \end{tabular}
   \caption{Relative signs of the contributions to the differential rate for $\tau^-$ and $\tau^+$ decays.}
   \label{tab:tauplus}
\end{table}

%%%%%%%%%%%%%%%%%%%%%%%%%%%%%%%%%%%%%%%%
\section{Feasibility study at Belle and Belle II} \label{sec:fstudy}
%%%%%%%%%%%%%%%%%%%%%%%%%%%%%%%%%%%%%%%%

In this section we outline our technique to estimate the sensitivity on $\tau$ dipole moments via $\tau$ leptonic radiative decays. First, however, we will discuss the possibility, suggested in ref.~\cite{Laursen:1983sm}, to determine $\tilde{a}_{\tau}$ taking advantage of the radiation zero which occurs in the radiative leptonic decays $\tau\to{l}\nu\nu\gamma$ for $c=-1$ (i.e., ${l}$ and $\gamma$ back-to-back in the $\tau$ rest frame) and maximal energy of the lepton ${l}$, i.e.\ $x^{\rm max}=2E^{\rm max}_{{l}}/m_{\tau}=1+r^2$. To this end, we analyzed a set of $\tau^+\tau^-$ events, where one $\tau$ decays to the radiative leptonic mode and the other $\tau$ decays to ordinary (inclusive) leptonic mode:
$\tau^{\pm}\to{l}^{\pm}_1\nu\nu\gamma,~\tau^{\mp}\to{l}^{\mp}_2\nu\nu$,
with ${l}_{1,2}=e$ or $\mu$, and ${l}_{1}\neq{l}_{2}$ --- in short: 
$({l}^{\pm}_1\gamma,~{l}^{\mp}_2)$. 
We excluded 
$(e^{\pm}\gamma,~e^{\mp})$ and $(\mu^{\pm}\gamma,~\mu^{\mp})$ 
events from our analysis because of the large background from 
$e^+ e^- \to e^+ e^- \gamma$ and $e^+ e^- \to \mu^+ \mu^- \gamma$ 
processes. The analyzed events were produced by the KKMC/TAUOLA/PHOTOS generators~\cite{Jadach:1999vf,Jadach:1993hs,Barberio:1993qi} and processed by GEANT3 based program~\cite{geant} in the conditions of the Belle experiment.

The sensitivity to $\tilde{a}_{\tau}$ is determined by the background suppression power $\varepsilon_{\rm sig}/\varepsilon_{\rm bg}$, where $\varepsilon_{\rm sig}$ is the detection efficiency for signal events and $\varepsilon_{\rm bg}$ is that for background events. 
The main background comes from the SM radiative leptonic decays (characterized by $\tilde{a}_{\tau}=0$) as well as from $(\tau^+\to{l}^+_1\nu\nu;~\tau^-\to{l}^-_2\nu\nu)\gamma_{\mysmall ISR}$ events with initial state radiation (ISR) towards large polar angles in the detector. 
As the fraction of the signal events in the vicinity of the radiation zero point is very small, we extended the signal 
region to maximize $\varepsilon_{\rm sig}/\varepsilon_{\rm bg}$: 
\begin{equation}
   0.1<\cos{\widehat{({l}_2,\gamma)}}<0.8, \quad
   \cos{\widehat{({l}_1,\gamma)}}<-0.9,\quad
   {\rm and}~E_{\gamma}>0.5~{\rm GeV}.
\end{equation}
Even in this case, the $\tilde{a}_{\tau}$ upper limit (UL) which can be achieved with the whole Belle statistics of about $0.9\times 10^9$ $\tau$ pairs is only UL$(\tilde{a}_{\tau})\simeq 2$. We found that the phenomenon of radiation zero 
has no large influence on the $\varepsilon_{\rm sig}/\varepsilon_{\rm bg}$. 
The dynamical structure of the signal events, determined by $G_a(x,y,c)$ (for this specific analysis, also terms of $O(\tilde{a}_{\tau}^2)$ were kept), allows us to achieve $\varepsilon_{\rm sig}/\varepsilon_{\rm bg}\sim 100$ only. At the same time, the suppression of the signal branching fraction for $\tilde{a}_{\tau}=1$ is ${\cal B}_{\rm bg}/{\cal B}_{\rm sig}\simeq 2000$, i.e.\ about one order of magnitude larger than $\varepsilon_{\rm sig}/\varepsilon_{\rm bg}$. As a result, there is no possibility to improve significantly the $\tilde{a}_{\tau}\sim 1$ sensitivity. Our feasibility study in the conditions of the Belle experiment therefore shows that the radiation zero method does not help to improve the present limits on $\tilde{a}_{\tau}$.

\begin{figure}[htb]
   \centering
   \includegraphics[width=0.6\textwidth]{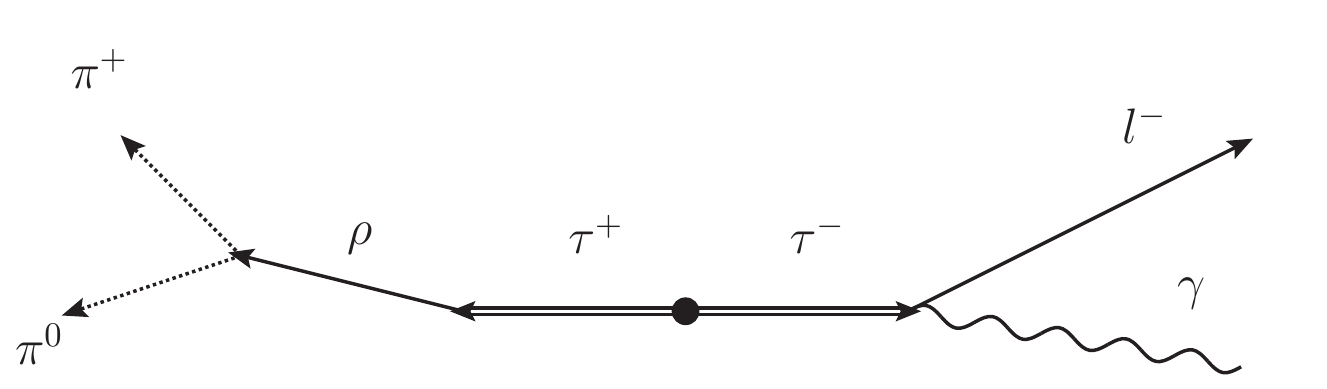}
   \caption{The $\rho$-tag mode used in the unbinned maximum likelihood fit. Events are analyzed in the $12$-dimensional phase space of $({l}^{\mp},\gamma,\pi^{\pm},\pi^0)$. Undetected neutrinos are not drawn.}
   \label{fig:rhotag}
\end{figure}

We will now outline our method to extract $\tilde{a}_{\tau}$ and $\tilde{d}_{\tau}$, which consists in the use of an unbinned maximum likelihood fit of events in the full phase space. The main idea is to consider events where both $\tau$ leptons decay to particular final states. One $\tau^{\mp}$ (signal side) decays to the radiative leptonic mode and the other $\tau^{\pm}$ (tag side) decays to some well-investigated mode with a large branching fraction. 
As a tag decay mode we choose $\tau^{\pm}\to\rho^{\pm}\nu\to\pi^{\pm}\pi^0\nu$ ($\rho$-tag mode), which also serves as spin analyser and allows us to be sensitive to the spin-dependent part of the differential decay rate of the signal decay using effects of spin-spin correlation of the $\tau$ leptons~\cite{Tsai:1971vv}. With this technique we analyzed $({l}^{\mp}\nu\nu\gamma,~\pi^{\pm}\pi^0\nu)$ events in the 12-dimensional phase space (PS), see figure~\ref{fig:rhotag}.

The probability density function (PDF) is constructed from the total differential cross section 
$\frac{d\sigma}{\rm dPS}(e^+ e^-\to \tau^{\mp}\tau^{\pm}\to ({l}^{\mp}\nu\nu\gamma,~\pi^{\pm}\pi^0\nu))$, 
which is given by the sum of a spin-independent term and spin-spin correlation term. To write the total differential cross section we followed the approach developed in refs.~\cite{Fetscher:1990su,Tamai:2003he}. The differential cross section of $e^+e^- \to \tau^+(\hat{n}^+) \, \tau^-(\hat{n}^-)$ in the center-of-mass system (c.m.s.) is given by~\cite{Tsai:1971vv} (asterisks indicate parameters measured in the c.m.s.): 
\begin{equation} 
   \frac{d\sigma(\hat{n}^-,\hat{n}^+)}{d\Omega^*_{\tau}} = 
   \frac{\alpha^2\beta^*_{\tau}}{64E^{*2}_{\tau}} \left[ D_0+D_{ij} \, n^-_i n^+_j \right],
   \label{eqtotdif}
\end{equation}
where 
   $D_0 = 1+\cos^2{\theta^*_\tau}+\sin^2{\theta^*_\tau}/\gamma^{*2}_{\tau}$,
\begin{equation}
D_{ij} = \left( \begin{array}{@{}c@{~~}c@{~~}c@{}}
(1+\frac{1}{\gamma^{*2}_{\tau}})\sin^2{\theta^*_\tau} & 0 & \frac{1}{\gamma^*_{\tau}}\sin{2\theta^*_\tau} \\
 0 & -\beta^{*2}_{\tau}\sin^2{\theta^*_\tau} & 0 \\ 
\frac{1}{\gamma^*_{\tau}}\sin{2\theta^*_\tau} & 0 & 1+\cos^2{\theta^*_\tau}-\frac{1}{\gamma^{*2}_{\tau}}\sin^2{\theta^*_\tau} \\ 
\end{array} \right),
\end{equation}
and $\hat{n}^\mp$ is the polarisation vector of $\tau^{\mp}$ in its rest frame (unit three-vector along the $\tau^{\mp}$ spin direction with components $n^\mp_i$). Moreover, $E^*_{\tau}$, $\gamma^*_{\tau}=E^*_{\tau}/m_{\tau}$, $\beta^*_{\tau}=|\vec{p}_{\tau}^{~*}|/E^*_{\tau}$ and $\theta^*_\tau$ are the energy, Lorentz factor, velocity of the $\tau$ and the polar angle of the $\tau^-$ three-momentum $\vec{p}_{\tau}^{~*}$, respectively. The signal differential decay width, discussed earlier in section~\ref{sec:effectiveltau}, can be written in the form (with an unimportant, for this analysis, total normalization constant $\kappa_{{l}\gamma}$): 
\begin{equation}
   \frac{d\Gamma(\tau^{\mp}(\hat{n}^{\mp})\to{l}^{\mp}\nu\nu\gamma)}
   {dx \, dy \, d\Omega_{{l}} \, d\Omega_{\gamma}}= \kappa_{{l}\gamma}
   \left[
      A(x, y, z)
      \pm\hat{n}^{\mp} \! \cdot \vec{B}^{\mp}(x, y, z)
   \right], 
\end{equation} 
where
\begin{align} 
A(x, y, z) =  & \,\, x\beta_{l}\biggl[G(x, y, c, y_0) + \tilde{a}_{\tau}G_a(x, y, z)\biggr] 
\\
\vec{B}^{\mp}(x, y, z)  = & \,\, x\beta_{l}\biggl[\hat{p}_{{l}}x\beta_{l} \left(J+\tilde{a}_{\tau}J_a\right) 
			     + \hat{p}_{\gamma} y \left(K+\tilde{a}_{\tau} K_a\right) + \biggr.
\\
& \biggl.	\,\, 
+ (\hat{p}_{{l}}\times\hat{p}_{\gamma}) x y \beta_{l}\left(\pm L + (m_{\tau}/e) \tilde{d}_{\tau} L_d\right)\biggr].
\end{align}

The $\tau^{\pm}(\hat{n}^{\pm}) \to \rho^{\pm}(K) \, \nu(q) \to \pi^{\pm}(p_1) \, \pi^0(p_2) \, \nu(q)$ differential 
decay rate is (with a total normalization constant $\kappa_{\rho}$): 
\begin{equation}
   \frac{d\Gamma(\tau^{\pm}(\hat{n}^{\pm})\to\pi^{\pm}\pi^0\nu)}
   {dm^2_{\pi\pi} \, d\Omega_{\rho} \, d\Omega_{\pi\rho}} = \kappa_{\rho}
   \left[A' \mp \hat{n}^{\pm} \!\cdot \vec{B'}\right] W(m^2_{\pi\pi}),
\end{equation}
where
\begin{align}
   & A'= 2 \, (q \cdot Q) \, Q_0-Q^2q_0, & 		&\vec{B'} =Q^2\vec{K} + 2\, (q \cdot Q) \, \vec{Q}, \notag
   \\
   & Q = p_1 - p_2, & 					&K = p_1 + p_2, \notag &
   \\
   & W(m^2_{\pi\pi}) = |F_{\pi}(m^2_{\pi\pi})|^2 \frac{|\vec{p}_{\rho}| |\vec{p}_{\pi\rho}|}{m_{\tau}m_{\pi\pi}}, 
   		& 							&m^2_{\pi\pi}=K^2, \notag &
   \\
   & |\vec{p}_{\rho}| = \frac{m_{\tau}}{2} \left(1-\frac{m^2_{\pi\pi}}{m^2_{\tau}}\right), &  
   	&|\vec{p}_{\pi\rho}| = \frac{\lambda^{\frac{1}{2}}(m^2_{\pi\pi},m^2_{\pi},m^2_{\pi^0})}{2m_{\pi\pi}}, 
\end{align} 
and $\lambda(x,y,z) \equiv x^2+y^2+z^2 -2xy-2xz-2yz$ is the K\"{a}llen function. Also, $\vec{p}_\rho$ and $\Omega_\rho$ are the three-momentum and solid angle of the $\rho$ meson in the $\tau$ rest frame, $\vec{p}_{\pi\rho}$ and $\Omega_{\pi\rho}$ are the three-momentum and solid angle of the charged pion in the $\rho$ rest frame, and $F_{\pi}(m^2_{\pi\pi})$ is the pion form factor with the CLEO parameterisation~\cite{Urheim:1997ag}. As a result, the total differential cross section for $({l}^{\mp}\gamma,\rho^{\pm})$ events can be written as \cite{Tsai:1971vv}:
\begin{equation}
   \frac{d\sigma({l}^{\mp}\gamma,\rho^{\pm})}
   {dE_{{l}}\,d\Omega_{{l}}\,dE_{\gamma}\,
   d\Omega_{\gamma}\,d\Omega_{\rho}\,dm^2_{\pi\pi}\,d\Omega_{\pi\rho}\,d\Omega^*_{\tau}} \,=\,
   \kappa_{{l}\gamma}\kappa_{\rho} \,
   \frac{\alpha^2\beta^*_{\tau}}{64E^{*2}_{\tau}}
   \left[ D_0AA'-D_{ij}B^{\mp}_i B'_j \right] W(m^2_{\pi\pi}). 
\end{equation}

\begin{figure}[htbp]
   \centering
   \includegraphics[width=0.5\textwidth]{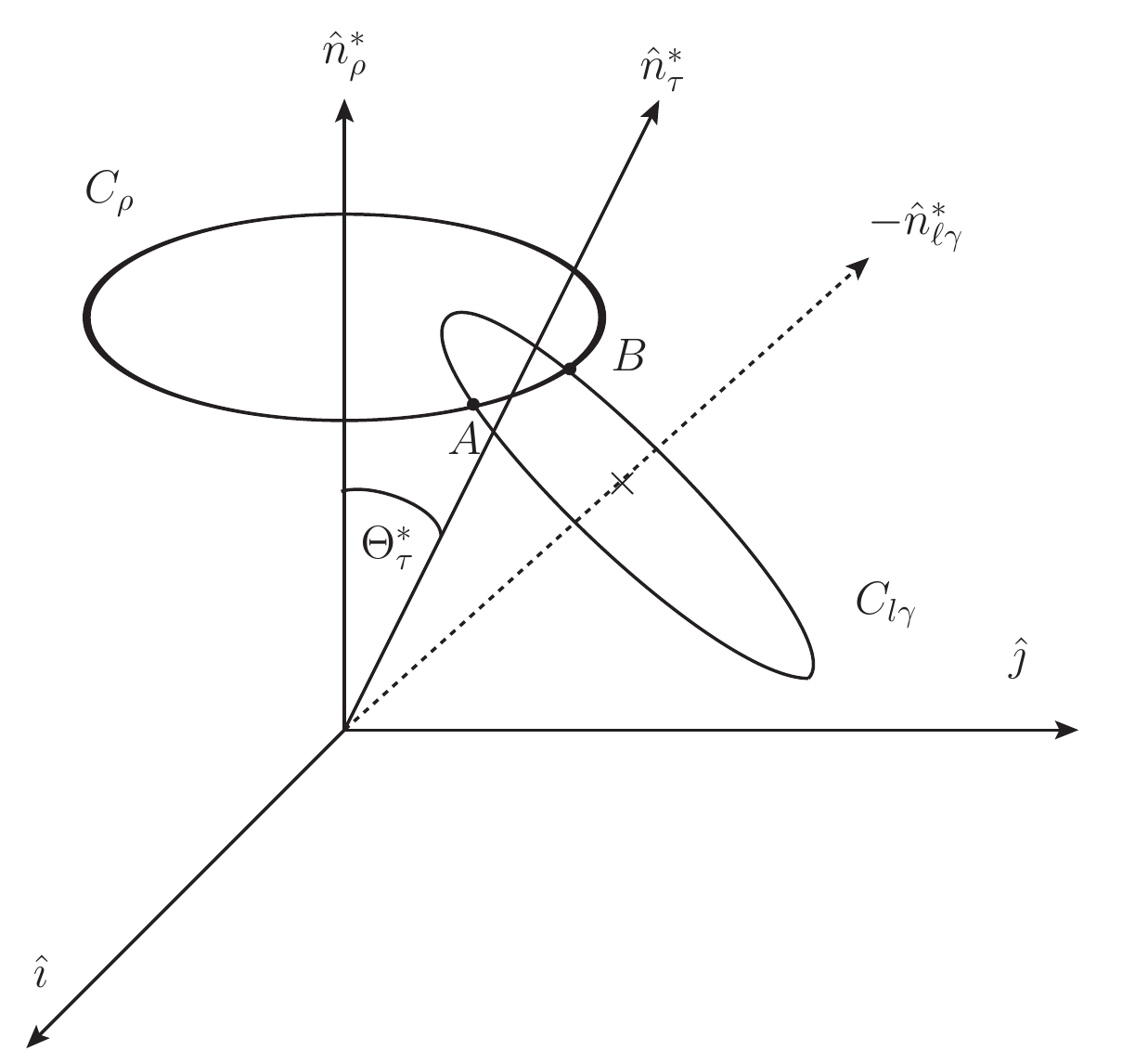}
   \caption{Configuration of the two circles $C_\rho$ and $C_{{l}\gamma}$ on a unit sphere, which are determined by the decays $\tau^+ \to \rho^+ \nu$ and $\tau^- \to {l}^- \nu \bar{\nu} \gamma$, respectively. The kinematically allowed $\tau$ direction in the c.m.s. is given by the intersection between the circumference of $C_\rho$ and spherical sector constrained by $C_{{l}\gamma}$.}
\label{fig:tausystem}
\end{figure}

In the c.m.s., the $\tau^\mp$ directions are limited on an arc $(\Phi^*_A,\Phi^*_B)$. The neutrino mass constraint in the decay $\tau^+ \to \rho^+ \nu$ gives the $\tau^+$ production angle, $\Theta^*_\tau$, with respect to the $\rho$ direction $\hat{n}^*_\rho$. This relation indicates that the $\tau^+$ direction $\hat{n}^*_\tau$, which lies on a unit sphere, is on the circumference of a circle $C_\rho$ with radius equal to $\sin \Theta^*_\tau$, as shown in figure~\ref{fig:tausystem}. Similarly, the invariant mass $m_{\nu \bar{\nu}}>0$ of the two-neutrino system in the decay $\tau^- \to {l}^- \nu \bar{\nu} \gamma$ gives a constraint on $\Theta^{*'}_\tau$, the $\tau$ angle along the direction of the ${l}\gamma$ system. The inequality $m_{\nu \bar{\nu}}>0$ confines the vector $\hat{n}^*_\tau$ to be either inside or outside the circle $C_{{l}\gamma}$, depending on the kinematics. Therefore, in the c.m.s., the direction of the $\tau^\mp$ system is given by the intersection between the circumference of $C_\rho$ and spherical sector constrained by $C_{{l}\gamma}$, i.e.\ the arc $(\Phi^*_A,\Phi^*_B)$.\footnote{We observed in the analysis that the constraint $m_{\nu\nu} < m_\tau - m_{l}$ did not provide additional information on the $\tau$ direction.}

Experimentally one measures particle parameters in the c.m.s. Therefore, defining 
$\vec{X}=(|\vec{p}_{l}^{~*}|,\Omega^*_{{l}},|\vec{p}_{\gamma}^{~*}|,\Omega^*_{\gamma},|\vec{p}_{\rho}^{~*}|,\Omega^*_{\rho},m^2_{\pi\pi},\Omega_{\pi\rho})$, 
the visible differential cross section is~\cite{Tamai:2003he}:
\begin{equation}
   {\cal F} (\vec{X}) = \frac{d\sigma({l}^{\mp}\gamma,\rho^{\pm})}{d \vec{X}}
   = \int_{\Phi^*_A}^{\Phi^*_B} \frac{d\sigma({l}^{\mp}\gamma,\rho^{\pm})}
   {dE_{l}d\Omega_{l}dE_{\gamma} d\Omega_{\gamma}d\Omega_{\rho}
   dm^2_{\pi\pi}d\Omega_{\pi\rho}d\Omega^*_{\tau}} \,\, J \,\, 
    d\Phi^*_{\tau}, 
\label{eqn:crosec}
\end{equation}
where the integration is done over the unknown $\tau$ direction, which is constrained to lie on the $(\Phi^*_A,\Phi^*_B)$ arc. Both angles $\Phi^*_A$ and $\Phi^*_B$ are calculated using parameters measured by the experiment. The Jacobian $J$ in eq.~\eqref{eqn:crosec} can be simplified as: 
\begin{equation} 
J = \left\vert \frac{\partial (E_{l},\Omega_{l},E_{\gamma},\Omega_{\gamma},\Omega_{\rho},\Omega^*_{\tau})}
   {\partial (|\vec{p}_{l}^{~*}|,\Omega^*_{l},|\vec{p}_{\gamma}^{~*}|,\Omega^*_{\gamma},|\vec{p}_{\rho}^{~*}|,\Omega^*_{\rho},\Phi^*_{\tau})}\right\vert = 
\biggl|\frac{\partial (E_l,\Omega_l)}{\partial (|\vec{p}_{l}^{~*}|,\Omega^*_{l})}\biggr| \,\, \biggl|\frac{\partial (E_{\gamma},\Omega_{\gamma})}{\partial (|\vec{p}_{\gamma}^{~*}|,\Omega^*_{\gamma})}\biggr| \,\, \biggl|\frac{\partial (\Omega_{\rho},\Omega^*_{\tau})}{\partial (|\vec{p}_{\rho}^{~*}|,\Omega^*_{\rho},\Phi^*_{\tau})}\biggr|,
\end{equation} 
where
\begin{align}
   & \biggl|\frac{\partial (E_{\alpha},\Omega_{\alpha})}{\partial (|\vec{p}_{\alpha}^{~*}|,\Omega^*_{\alpha})}\biggr| = 
\frac{|\vec{p}_{\alpha}^{~*}|^2}{E^*_{\alpha} |\vec{p}_{\alpha}|}, 
\quad \mbox{with }  \alpha = l, \gamma,\\
& \biggl|\frac{\partial (\Omega_{\rho},\Omega^*_{\tau})}{\partial (|\vec{p}_{\rho}^{~*}|,\Omega^*_{\rho},\Phi^*_{\tau})}\biggr| = \frac{m_\tau}{|\vec{p}_{\tau}^{~*}|}\frac{|\vec{p}_{\rho}^{~*}|}{E^*_\rho |\vec{p}_{\rho}|}.
\end{align}
In our feasibility study we developed a special generator of the signal $({l}^{\mp}\nu\nu\gamma,~\pi^{\pm}\pi^0\nu)$ events. For the unbinned maximum likelihood fit of the generated events, the PDF is constructed as:
\begin{equation} 
{\cal P}(\vec{X})=\frac{{\cal F}(\vec{X})}{\int{\!\cal F}(\vec{X}) \, d\vec{X}}.
\end{equation}
Fitting samples of generated events corresponding to the amount of data available at Belle and expected at Belle II, we studied the sensitivities to the parameters $\tilde{a}_{\tau}$ and $\tilde{d}_{\tau}$.

Our results are collected in table~\ref{tabres}, where the sensitivities are shown for two cases: ({\textit i}) events are tagged by $\tau^{\pm}\to\rho^{\pm}\nu$ only ($\rho$-tag); (\textit{ii}) six decay modes 
($\tau^{\pm}\to\rho^{\pm}\nu$, $\tau^{\pm}\to\pi^{\pm}\nu$, 
$\tau^{\pm}\to\pi^{\pm}\pi^0\pi^0\nu$, $\tau^{\pm}\to\pi^{\pm}\pi^+\pi^-\nu$, 
$\tau^{\pm}\to e^{\pm}\nu\nu$, $\tau^{\pm}\to\mu^{\pm}\nu\nu$)  with a total branching fraction of about $90\%$  are used for the tag (full tag). In the full-tag case, the sensitivity increase is due to the statistical factor $\sqrt{90/25.5}=1.88$, compared to the $\rho$-tag case with ${\cal B}=25.5\%$.
We note that the integration over the arc $(\Phi^*_A,\Phi^*_B)$ inflates the uncertainty by a factor of $1.4$ in comparison with the case when the direction of the $\tau$ is known. Also, the inclusion of the spin-dependent part of the differential decay rate increases the sensitivity by a factor of about 1.5. It is interesting to note that the sensitivity for events with $\tau \to e \nu \bar{\nu} \gamma$ is two times worse than that for $\tau \to \mu \nu \bar{\nu} \gamma$ (with the same statistics).
Table~\ref{tabres} also shows, for comparison, the sensitivities to $\tilde{a}_{\tau}$ and $\tilde{d}_{\tau}$ obtained in the most precise previous studies at DELPHI~\cite{Abdallah:2003xd} and Belle~\cite{Inami:2002ah}, respectively. It can be clearly seen that the measurement of $\tilde{a}_{\tau}$ in $\tau$ radiative leptonic decays at Belle II with the full tag can improve the DELPHI result. On the other hand, the expected sensitivity to $\tilde{d}_{\tau}$ is still worse than the most precise measurement of $\tilde{d}_{\tau}$ performed at Belle in $\tau^+\tau^-$ pair production. 
\begin{table}[htbp]
\centering
\caption{Sensitivities to $\tilde{a}_{\tau}$ and $\tilde{d}_{\tau}$ in $\tau$ radiative leptonic decays ($\rho$-tag and full-tag cases) which can be achieved with the whole data sample collected at Belle and planned for Belle~II. The present most precise results by DELPHI~\cite{Abdallah:2003xd} and Belle~\cite{Inami:2002ah} are shown in the last two columns. $(m_{\tau}/e)=9.0 \times 10^{13}(e  \mathrm{\cdot cm})^{-1}.$}
\label{tabres}

\begin{tabular}{c|llllll}
\toprule
						& Belle ($\rho$)	&  Belle~II ($\rho$)	& Belle (full)	& Belle~II (full) 											& DELPHI~\cite{Abdallah:2003xd}	& Belle~\cite{Inami:2002ah} \\ 
\midrule
$\tilde{a}_{\tau}$   			&  $0.16$			&  $0.023$			& 0.085 			& 0.012									& 0.017						& --- \\
\midrule
$(m_{\tau}/e)\,\tilde{d}_{\tau}$	&  $0.15$			&  $0.021$			& 0.080 			& 0.011									& --- 							& 0.0015 \\                
\bottomrule
\end{tabular}
\end{table}
% 

%%%%%%%%%%%%%%%%%%%%%%%%%%%%%%%%%%%%%%%%
\section{Conclusions}\label{sec:conclusions}
%%%%%%%%%%%%%%%%%%%%%%%%%%%%%%%%%%%%%%%%

The magnetic and electric dipole moments of the $\tau$ lepton are largely unknown. Several proposals have been presented in the past to study them, but the current sensitivity is only of $\mathcal{O}(10^{-2})$ for $a_{\tau}$ and $\mathcal{O}(10^{-3})$ for $d_{\tau}$. In this article we presented a new method to probe $a_{\tau}$ and $d_{\tau}$ using precise measurements of the differential rates of radiative leptonic $\tau$ decays at high-luminosity $B$ factories. In our approach, deviations of the $\tau$ dipole moments from the SM predictions are determined via an effective Lagrangian, thus yielding model-independent results. To this end, in section~\ref{sec:effectiveltau} we provided explicit analytic formulae for the relevant non-standard contributions to the differential decay rates generated by the effective operators contributing to the $\tau$ \gmt~and EDM. These expressions, combined with the SM predictions recently computed at NLO in~\cite{Fael:2015gua}, can be compared with precise data to probe the $\tau$ dipole moments. Earlier proposals to determine the $\tau$ anomalous magnetic moment were examined in sections~\ref{sec:status} and~\ref{sec:fstudy}.

Our technique to estimate the sensitivity on $\tau$ dipole moments via $\tau$ leptonic radiative decays was outlined in section~\ref{sec:fstudy}, where we presented a detailed feasibility study of our method in the conditions of the Belle and (upcoming) Belle~II experiments. The results of this study are summarized in table~\ref{tabres}. They show that our approach, applied to the planned full set of Belle~II data for radiative leptonic $\tau$ decays, has the potential to improve the present experimental bound on the $\tau$ \gmt. On the contrary, the foreseen sensitivity is not expected to lower the current experimental limit on the $\tau$ EDM.

%%%%%%%%%%%%%%%%%%%%%%%%%%%%%%%%%%%%%%%
\acknowledgments
%%%%%%%%%%%%%%%%%%%%%%%%%%%%%%%%%%%%%%%
We would like to thank A.~Crivellin, S.\ Rigolin, A.~Santamaria and Z.~Was for very useful discussions and correspondence. 
S.E.\ and D.E.\ thank Prof.~H.~Aihara, C.~Ng and F.~Okazawa (University of Tokyo) for fruitful discussions and great help in the development of the necessary software. 
The work of M.F.\ is supported by the Swiss National Science Foundation.
M.P.\ also thanks the Department of Physics and Astronomy of the University of Padova for its support. His work was supported in part by the Italian Ministero dell'Universit\`a e della Ricerca Scientifica under the program PRIN 2010-11, and by the European Program INVISIBLES (contract PITN-GA-2011-289442).

%%%%%%%%%%%%%%%%%%%%%%%%%%%%%%%%%%%%%%%%
\bibliographystyle{JHEP}
  \footnotesize
\bibliography{Bibliography}

\end{document}